\input{aipcheck.tex}

\newcommand\sps{\space\space\space\space}
  \def\selectedoptions{final}

\documentclass[
   \selectedoptions
  ]
  {aipproc}

\layoutstyle{6x9}

\SetInternalRegister\hbadness{8000} 

\newcommand\doingARLO[2][]{%
  \ifx\mmref\undefined #1\else #2\fi
}

\begin{document}

\title 
      [GMRT Reionization]
      {
The GMRT Search for Reionization
}

\classification{}
\keywords{cosmology, reionization}

\author{Ue-Li Pen}{
  address={CITA, 60 St George St, Toronto, Canada, M5S 3H8},
  email={pen@cita.utoronto.ca},
  thanks={This research was support by NSERC}
}

\iftrue
\author{Tzu-Ching Chang}{
  address={Institute for Astronomy and Astrophysics, Academia Sinica,
  P.O. Box 23-141, Taipei 10617, Taiwan},
  address={CITA, 60 St George St, Toronto, Canada, M5S 3H8},
  email={tchang@cita.utoronto.ca},
}

\author{Jeff B. Peterson}{
  address={Department of Physics, Carnegie Mellon University, 5000 Forbes Ave, Pittsburgh, PA 15213, USA},
  email={jbp@cmu.edu},
}

\author{Jayanta Roy}{
  address={National Center for Radio Astrophysics, Tata Institute for 
Fundamental Research, Pune, India},
  email={jroy@ncra.tifr.res.in},
}

\author{Yashwant Gupta}{
  address={National Center for Radio Astrophysics, Tata Institute for 
Fundamental Research, Pune, India},
  email={ygupta@ncra.tifr.res.in},
}
\author{Kevin Bandura}{
  address={Department of Physics, Carnegie Mellon University, 500
    Forbes Ave, Pittsburgh, PA 15213, USA},  
  email={kbandura@andrew.cmu.edu},
}

\fi

\copyrightyear  {2008}

\begin{abstract}
We present an overview for the reionization search at GMRT.  The
forecast sensitivities are promising for an early detection.  RFI
mitigation has been successful.  Several hundred hours of telescope
time have already been invested in this ongoing effort, and analysis
of the data is in progress.
\end{abstract}

\date{\today}

\maketitle

\section{Introduction}

\def\lesssim{\mathrel{\hbox{\rlap{\hbox{\lower4pt\hbox{$\sim$}}}\hbox{$<$}}}}

A current frontier in observational and theoretical astrophysics is the
search for  structures during the epoch of reionization (EoR).
The WMAP satellite has measured polarization in the Cosmic
Microwave Background (CMB) at large angular scales.  This polarization
is believed to arise from Thomson scattering of the CMB photons near
the EoR \citep{2008arXiv0803.0593N,2008arXiv0803.0547K}. The observed
optical depth $z\sim 0.089\pm 0.016$ corresponds to an instantaneous
reionization redshift of $z_{\rm reion}=10.8\pm 1.4$.  The redshifted
neutral hydrogen 21cm line is accessible to existing radio telescopes.

One way to study the reionization transition is by imaging redshifted
21cm emission. At redshifts above the EoR transition the gas is
neutral and is predicted to glow with about 25 mK sky brightness
temperature. After reionization is complete this glow is absent.
At redshifts close to the transition a patchy sky is expected.
Simulations \citep{2008MNRAS.384..863I} suggest that with existing
telescopes a measurement near 150 MHz may allow for statistical detection
of $\sim$ 20 Mpc patchiness in the neutral hydrogen.  This detection
would pin down the reionization redshift and begin the process of a more
detailed study of the transition.

Several programs\footnote{Giant Metrewave Radio Telescope (GMRT;
{\url{http://www.ncra.tifr.res.in}}), Low Frequency Array (LOFAR;
  \url{http://www.lofar.org}), Murchison Widefield Array (MWA;
  {\url{http://web.haystack.mit.edu/arrays/MWA}}), Primeval Structure
  Telescope/21CMA (PAST; \url{http://web.phys.cmu.edu/~past/}),
  and Square Kilometre Array (SKA; {\url{http://www.skatelescope.org}}).}
  are  underway to measure the collective 21cm emission during the
  epoch of reionization, $z>6$, when the neutral fraction of the universe
  was closer to unity. At the low frequency of these observations the
  continuum sources are very bright, so under these programs effective
  tools for continuum removal are being developed.

\section{Forecast}

\begin{figure}
 {\includegraphics[width=\textwidth]{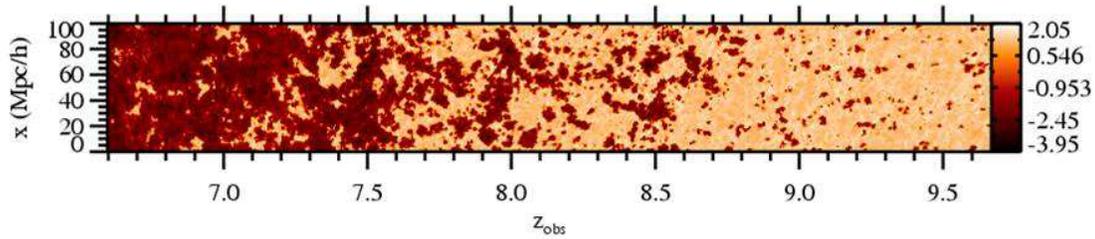}}
\caption{
A simulation \citep{2008MNRAS.384..863I} of the 21cm brightness field.  
The colour scale is in log milli
Kelvin.  The vertical scale is just under one degree.  Redshifts in
the range $6\lesssim z \lesssim 10$ are accessible at GMRT, with some
windows that are filtered for radio interference.  Reprinted with permission.
}
\label{fig:reionization}
\end{figure}

Over the past five years, the CITA group has invested in
realistic forecasts for reionization scenarios.  This group has
performed large scale radiative transfer reionization
simulations \citep{2008MNRAS.384..863I}, which allows for
quantitative forecasts of signals visible to GMRT. Figure \ref{fig:reionization}
shows how the universe might look through a slice visible to GMRT.

These simulations found that reionization power is present on large
(20 Mpc) scales, and that the optimal redshift to search for spatial
structure is at redshifts slightly less than the equivalent
instantaneous reionization redshift.

\begin{figure}
 {\includegraphics[height=.5\textheight]{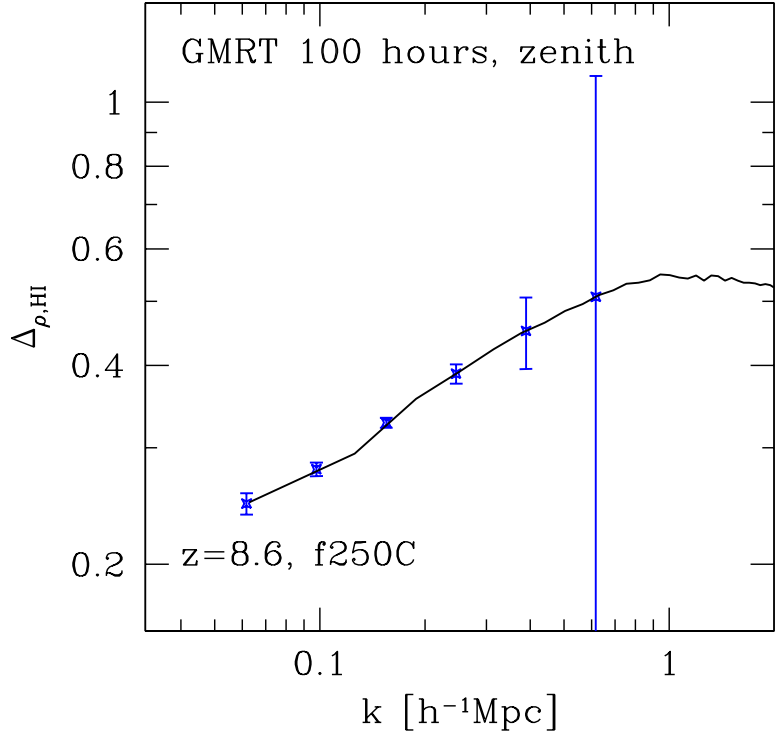}}
  \caption{
Observability of the 21-cm signal: the 3D power spectrum of the
neutral hydrogen density, $\Delta_{\rho,HI}$, at redshift $z=8.59$
(where the mean spin temperature $\overline{T_{b}}=16.3$~mK) with the
forecast error bars for 100 hours 
observation with GMRT vs. wavenumber $k$. We assumed 15 MHz observing
bandwidth (the full utilized 150 MHz bandwidth of GMRT), $T_{\rm
  sys}=480$~K and $T_S \gg T_{\rm CMB}$. The array configuration is
assumed pointed to 
the zenith, but the sensitivity is only weakly dependent on the pointing.
}
\label{fig:forecast}
\end{figure}

Estimated noise levels can be comparable to the signal per pixel on
scales down to 20 Mpc in a 100 hour observation.  Maps will be noisy,
but the power spectrum may still be accurately measurable.  Figure
\ref{fig:forecast} shows the expected error on a reconstructed power
spectrum in a 100 hour observation, assuming that thermal limits are
reached, and that all foregrounds have been subtracted.

\section{Telescope}

Our group has initiated an effort to search for 21cm structures at the
epoch of reionization using the Giant Metrewave Radio Telescope (GMRT)
in India.  This effort began in the summer of 2005, through a series
of visits and agreements.  At the time, a major challenge in the 2m
band was broad band interference, which was thought to be a
significant limitation for imaging dynamic range, especially on the
short base lines of interest for reionization studies.

\begin{figure}
{\includegraphics[width=\textwidth]{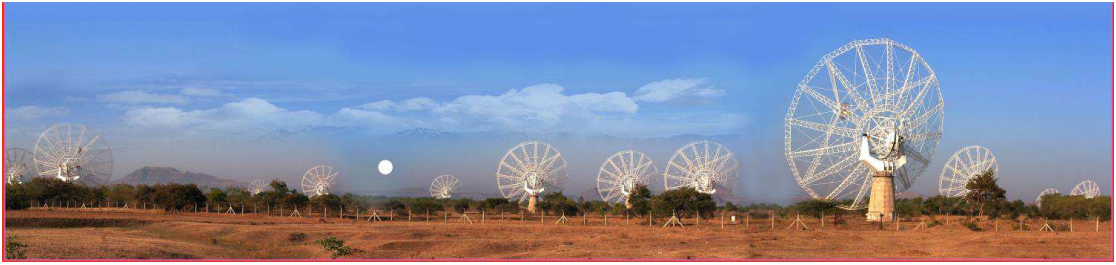}}
\caption{GMRT central square antennae.  The closely spaced antennae
  provide the high brightness sensitivity to search for EoR.}
\label{fig:gmrt}
\end{figure}

The telescope consists of 30 antennae of 45m diameter each.  14 are
designated ``central core'' antennae, which are within a 1 km area.
Figure \ref{fig:gmrt} shows a view of several central core antennae.
The dense layout of the core allows high brightness sensitivity, which
is needed for the search for reionization.  For this experiment, the
150 MHz feeds were used.  These consist of orthogonal pairs of folded
dipoles, backed by a ground plane.  These antennas couple to X and Y
linear polarizations, but the two signals pass through a hybrid
coupler before entering the amplifiers.  For each antenna this results
in a pair of right and left circularly polarized signals which are
amplified, up-converted and transmitted optically to the receiver
room.  Later processing allows measurement of the full set of Stokes
parameters.

\begin{figure}
  {\includegraphics[width=.5\textwidth]{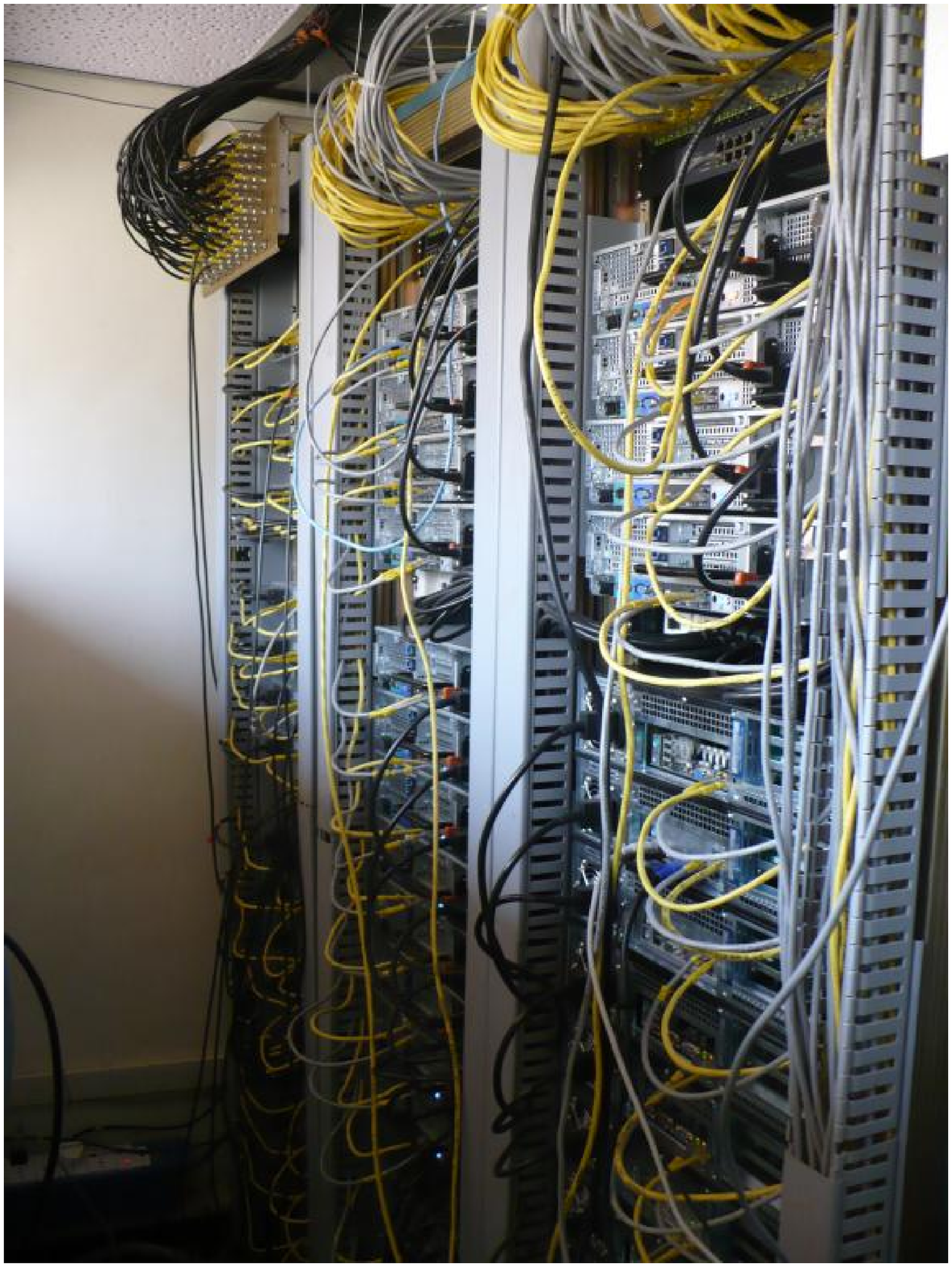}}
  {\includegraphics[width=.5\textwidth]{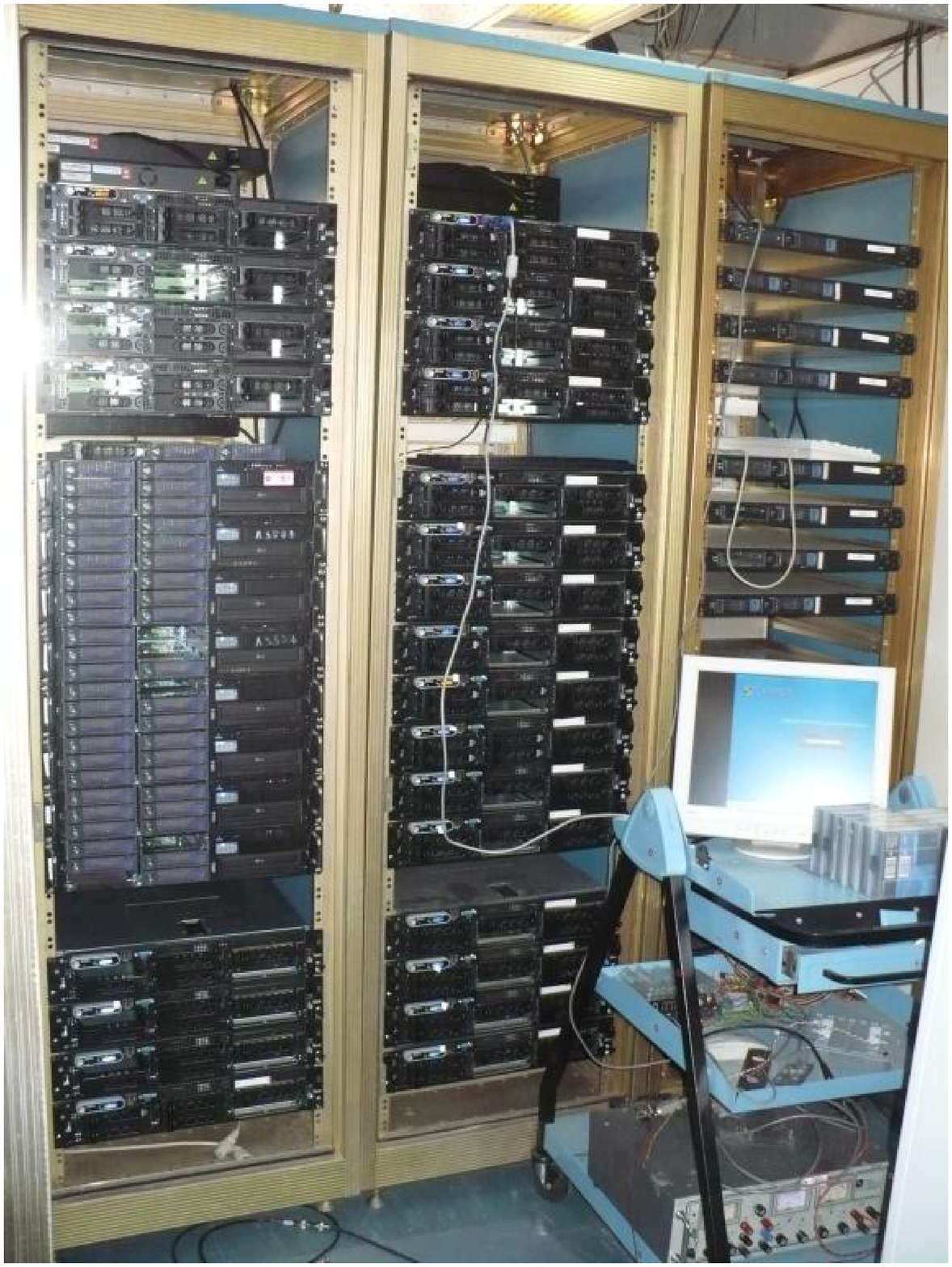}}
  \caption{The newly installed GMRT software back end.  It allows a
  wide variety of modes, including software correlation, and recording
  of all raw voltages to disk.}
\label{fig:gsb}
\end{figure}

To attempt the EoR experiment, and for other work at GMRT, we built a
new signal processing system for the telescope.  This consists of 16
commercial Analog to Digital sampling boards installed in an array of
off-the-shelf computers, as shown in Figure \ref{fig:gsb}, called the
GMRT Software Backend (GSB, Roy et al 2008, in prep).  The AD boards
have 4 input channels, and are connected to a common clock and trigger
signal, allowing synchronous sampling.  The GSB is in principle a
fully flexible software signal processing system, and can be
arbitrarily reconfigured in software.  The codes are written in C with
vendor library calls for FFT and inline vector assembly directives to
achieve close to theoretical peak speed of 4.77 Tera 16 bit fixed
point operations/sec on 16 nodes of dual quad core 2.33 GHz
processors.  The data is interchanged through dual channel gigabit
ethernet.  Initially, the GSB allows three modes of operation: fully
real time correlations, gated observations for pulsar and power line
RFI folding, and a raw recording mode for off-line processing.

For this experiment, only 16 MHz out of potentially possible width of
32 MHz of RF bandwidth are used.  The bandwidth was throttled by an IF
filter.  All 60 signals were sampled with 8 bit precision at 33
MSample/s.  Each AD board transfers the digitized data into its
individual host computer.  These streams are Fourier transformed in
blocks of length 4096 samples at 16 bit precision.  The 2048 complex
Fourier coefficients are then rescaled to 4 bits, and sent over a
gigabit network to correlation nodes.  Each block of 128 frequencies
is sent to one of 16 software correlation nodes.  These products,
which we call visibilities, are accumulated for 1/4 second in 16
distinct gates.  Thus the initial visibilities have a frequency and
time resolution of 7.8 kHz and 1/4 second (without counting for the
gates), respectively.

To simplify calibration, fields are taken along the lines of sights
through pulsars.  The ``gates'' are essential to our polarization
calibration technique.  These gates in time are synchronized so each
covers one of 16 segments of the pulsation cycle of a pulsar in the
field. The pulsar is a known source of polarized emission. By
comparing pulsar-on to pulsar-off we can measure the system gain and
phase directly using a sky source.  The raw 1/4 second averaged
visibilities are stored on disk, as well as 16 fringe-stopped gated
visibilities integrated for 16 seconds, and averaged over frequency
into 128 frequency channels.  The GSB allows allows decoding of the
GMRT noise source injection system, which enables a direct measurement
of the variable pulsar amplitude.

All these signal processing calculations occur simultaneously, and are
structured as individual asynchronous pipelined processes.  The processor
and network speeds are sufficient that each calculation is completed in
real time and there is less than 10 \% data loss.

\section{Interference Removal}

\begin{figure}
  {\includegraphics[height=.5\textheight]{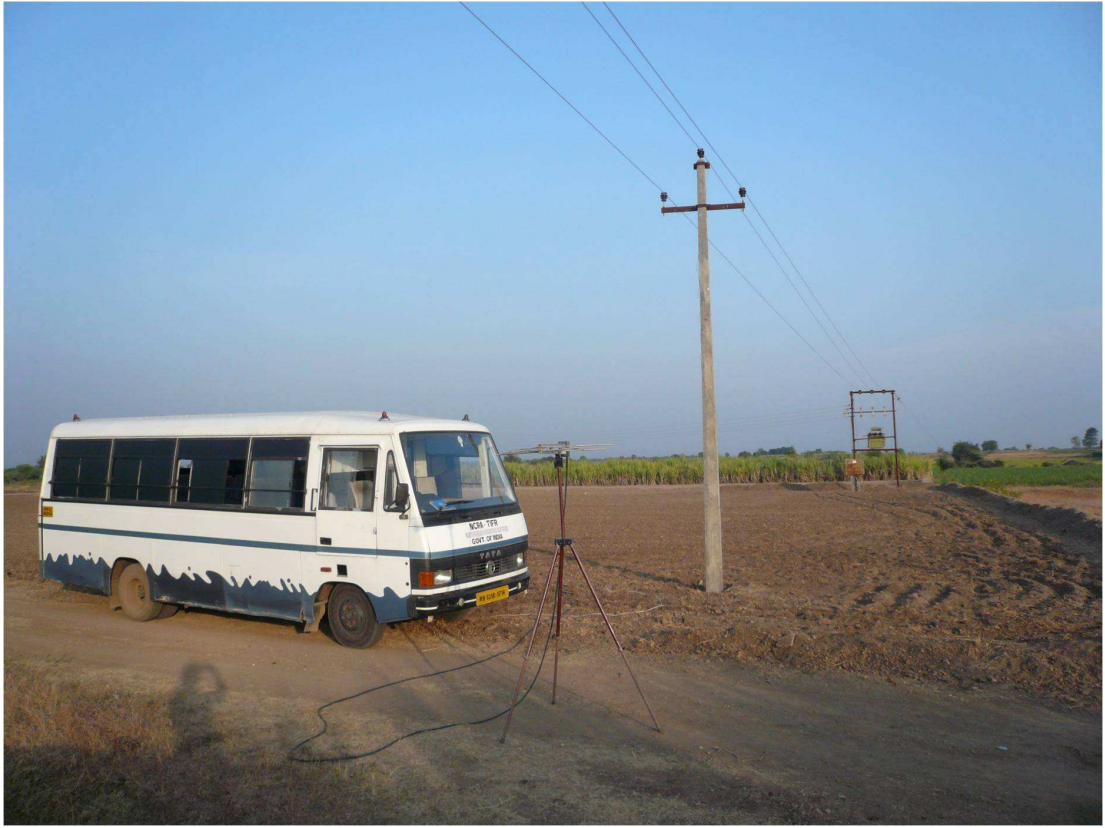}}
  \caption{RFI localization setup.  Power lines are thought to be
the primary cause of broadband interference.}
\label{fig:rfi}
\end{figure}

In the EoR project, a two pronged effort has been implemented to
address broad band interference problems.  The first is a software
processing step.  The second, which is still under development, is an
effort to localize the sources physically based on their inferred near
field position.  New near field calibration and localization
algorithms have been implemented and tested for this purpose (Pen et
al 2008, in prep).  Figure \ref{fig:rfi} shows the calibration source
setup, and a targeted RFI source, probably the transformer, in the
background,

Interference is removed in two stages.  First, line RFI is flagged.
In each frequency bin, the distribution of intensities are calculated.
The upper and lower quartile boundaries are used to estimate the
amplitude of the noise.  Any outliers further than $3\sigma$ on a
Gaussian scale are masked.  Each mask is 8 kHz wide, and 4 seconds
long.  Approximately 1\% of the data is flagged by this process.

A bigger problem at GMRT has been broad band interference.  These are
particularly problematic at short baselines.  A singular value
decomposition (SVD) based approach was used.  The data can be
considered as a matrix.  Each hour scan has 14400 time records, which
we call the rows of our matrix, and each record has 2048x60x61 entries
that correspond to the number of frequency channels and baselines in
complex visibilities. This 14400x7495680 matrix is then SVD
decomposed.  The first 100 right eigenvectors are tagged as "noise",
and removed.  This removes 0.7\% of the information.  The rationale is
that sources on the sky fringe rotate, and do not factor easily into
vectors with large eigenvalues, while sources on the ground modulate
coherently with time.  Empirically, this works very well, and no broad
band RFI is visible in the data after this process.  

\section{Current Status}

The new signal processing system has been fully installed and functioning
since August 2007; in December 2007, we acquired 100 hours of EoR data
suitable for the pulsar-calibration technique.  Initial analysis shows
that this new technique works well and the pulsar-only data is able to
achieve thermal noise limits. The EoR analysis is currently underway.

\section{Conclusions}

The GMRT effort is well underway in for a sensitive search for
reionization in redshifted 21cm structures.  Significant progress has
been achieved in interference removal, calibration, and forecasting.
Data acquisition is in progress, and the forecast sensitivities may
deliver early reionization detections.

\begin{theacknowledgments}

We would like to thank the National Center for Radio Astronomy (NCRA)
in India for persistent support through telescope time,
infrastructure, and technical support.  Figures
\ref{fig:reionization}, \ref{fig:forecast} are reprinted with permission
from Wiley-Blackwell Publishing.  They were originally published by
\cite{2008MNRAS.384..863I}, with title ``Current models of the
observable consequences of cosmic reionization and their
detectability''.

The work is funded in part by NSERC.

\end{theacknowledgments}


\doingARLO[\bibliographystyle{aipproc}]
          {\ifthenelse{\equal{\AIPcitestyleselect}{num}}
             {\bibliographystyle{arlonum}}
             {\bibliographystyle{arlobib}}
          }
\bibliography{sample}

\end{document}